\documentclass[12pt,preprint,a4paper]{aastex}
\usepackage{times}

\usepackage{graphics,epsf}

\def \km{~\rm{km}}

\def \erg{~\rm{erg}}

\def \Kep{\mathrm{Kep}}
\def \zams{\mathrm{ZAMS}}
\def \nicefrac#1#2{{{#1}/{#2}}}

\begin{document}

\title{Triggering jet-driven explosions of core-collapse supernovae by accretion from convective regions}

\author{Avishai Gilkis\altaffilmark{1} and Noam Soker\altaffilmark{1}}

\altaffiltext{1}{Department of Physics, Technion -- Israel Institute of Technology, Haifa
32000, Israel; agilkis@tx.technion.ac.il; soker@physics.technion.ac.il}

\begin{abstract}
We find that convective regions of collapsing massive stellar cores possess sufficient stochastic angular momentum to form intermittent accretion disks around the newly born neutron star (NS) or black hole (BH), as required by the jittering-jets model for core-collapse supernova (CCSN) explosions. To reach this conclusion we derive an approximate expression for stochastic specific angular momentum in convection layers of stars, and using the mixing-length theory apply it to four stellar models at core-collapse epoch. In all models, evolved using the stellar evolution code MESA, the convective helium layer has sufficient angular momentum to form an accretion disk. The mass available for disk formation around the NS or BH is $0.1-1.2 M_\odot$; stochastic accretion of this mass can form intermittent accretion disks that launch jets powerful enough to explode the star according to the jittering-jets model. Our results imply that even if no explosion occurs after accretion of the inner $\sim 2-5 M_\odot$ of the core onto the NS or BH (the mass depends on the stellar model), accretion of outer layers of the core will eventually lead to an energetic supernova explosion.
\end{abstract}

\section{INTRODUCTION}
\label{sec:intro}

Massive star cores collapse upon reaching a critical mass of iron-group elements (``iron''). A massive ``iron'' core implies the presence of several convective burning shells, such as silicon and oxygen. If a star still holds part of its hydrogen-rich envelope, this envelope is also convective. The convective elements (cells) have stochastic velocity, hence stochastic angular momentum, and accretion of convective zones onto the newly formed neutron star (NS) may form intermittent accretion disks with stochastic angular momentum. Each such temporary accretion disk can then launch two opposite jets. These `jittering-jets' might explode the star without any additional neutrino energy \citep{Soker2010, Papish2011, Papish2012b, Papish2014a, Papish2014b}.

Interest in jet-driven core-collapse supernova (CCSN) explosion mechanisms has grown in recent years following the failure of neutrino-driven explosion studies to reach a consistent and persistent explosion. Over the years, neutrino-driven explosions \citep{Colgate1966}, mainly the so called  delayed-neutrino mechanism \citep{Wilson1985, bethe1985}, were the most well-studied, with research groups running sophisticated multidimensional hydrodynamical simulations with ever increasing capabilities (e.g. \citealt{bethe1985,Burrows1985,Burrows1995,Fryer2002,Ott2008,Marek2009,Nordhaus2010,Kuroda2012,Hanke2012,Bruenn2013,CouchOtt2013}). In many cases the simulations have failed to even revive the shock of the falling core material. Recent results have shown that in more realistic 3D numerical simulations explosions are even harder to achieve \citep{Janka2013,Couch2013,Takiwaki2013,Hanke2012,Hanke2013}. \cite{Nordhaus2010} and \cite{Dolence2013} on the other hand found it is easier to achieve shock revival in 3D simulations. \cite{Papish2012a} and \cite{Papishetal2014} argued that the delayed-neutrino mechanism has a generic character preventing it from exploding the star with the observed typical energy of $\sim 10^{51} \erg$. Even with shock revival, the typical explosion energy most numerical simulations achieve is only about $10^{50} \erg$, e.g., \cite{Suwa2013}. The problems of the delayed-neutrino mechanism can be overcome by a strong wind, either from an accretion disk \citep{kohri2005} or from the newly born NS. Indeed, where explosions with energies of $\sim 10^{51} \erg$ were achieved, it seems the driving force was a continuous wind from the center (e.g., \citealt{Bruenn2013}; see discussion in \citealt{Papish2014a}). Such a wind is not part of the delayed-neutrino mechanism, and most researchers consider it to have a limited contribution.

Among the different alternative explosion mechanisms \citep{Janka2012}, the most well studied are magnetohydrodynamics/jet-driven models (e.g. \citealt{LeBlanc1970, Meier1976, Bisnovatyi1976, Khokhlov1999, MacFadyen2001,Hoflich2001, Woosley2005, Burrows2007, Couch2009,Couch2011,Lazzati2011}). Most of these MHD models require a rapidly spinning core before collapse starts, and hence are only applicable for special cases. Recent observations (e.g. \citealt{Milisavljevic2013,Lopez2013,Ellerbroek2013}) show that jets might have a much more general role in CCSNe than suggested by these models.

\cite{Soker2010} and \cite{Papish2011} proposed a general jets-driven mechanism that in principle can explode all CCSNe. The sources of the angular momentum for disk formation are (1) instabilities in the shocked region of the collapsing core, e.g., the standing accretion shock instability (SASI), and (2) the convective regions in the core. Recent 3D numerical simulations show indeed that the SASI is well developed in the first second after core bounce \citep{Hanke2013} and the unstable spiral modes can amplify magnetic fields \citep{Endeve2012}. The spiral modes with the amplification of magnetic fields build the ingredients necessary for jet launching.

The latter angular momentum source, that of accretion from convective parts of the core and envelope, is the subject of this paper. In section \ref{sec:angularmomentum} we derive the expression for the specific angular momentum of matter accreted from a convective zone. In section \ref{sec:models} we apply the results to several stellar models and conclude that intermittent disks are likely to be formed. In the present study we emphasize the accretion of outer convective zones, i.e., oxygen shell and outward, to examine the prediction for very low energy SNe from neutrino mass loss during the formation of the NS \citep{Nadezhin1980,Lovegrove2013,Piro2013}. In section \ref{sec:implicaitons} we argue that such transient events are unlikely to be formed and summarize our main results.

\section{STOCHASTIC ANGULAR MOMENTUM IN CONVECTIVE ZONES}
\label{sec:angularmomentum}

We assume that a shell consists of convection elements with equal mass and size, and a random velocity $\overrightarrow{v}=v_{c}\left(\sin\theta\cos\varphi,\sin\theta\sin\varphi,\cos\theta\right)$
with a uniform probability density in $\theta$ and $\varphi$, where $v_{c}$ is the convection speed. The contribution of one element to the angular momentum in the (arbitrary) z direction is $J_{z}=\left(\overrightarrow{r}\times\overrightarrow{p}\right)\cdot\hat{z}$. The expectation value of $J_{z}$ is zero, but not the variance, which is
\begin{equation}
\mathrm{Var}(J_{z})=\left\langle J_{z}^{2}\right\rangle =(mv_{c}r)^{2}\frac{\int\left[\left(\hat{r}\times\hat{v}(\theta)\right)\cdot\hat{z}\right]^{2}d\Omega}{\int d\Omega}=\frac{1}{3}(mv_{c}r)^{2}\sin^{2}\theta_{r},
\label{eqvarj}
\end{equation}
where $\theta_{r}$ is the angle between the element's location and the $z$ axis. 
For a thin shell composed of N elements distributed isotropically we have
\begin{equation}
\sum\limits_{N} \sin^2\theta_{r} \rightarrow N\frac{\int \sin^2\theta_{r} d\Omega}{\int d\Omega} = \frac{2N}{3}.
\label{eqaveragetheta}
\end{equation}
The standard deviation of the \emph{specific} angular momentum for a thin shell is then
\begin{equation}
\sigma(j_{z})=\frac{\sqrt{\left\langle\mathrm{Var}(J_{z})\right\rangle}}{Nm}=\sqrt{\frac{2}{9N}}v_{c}r.
\label{eqsigmaj}
\end{equation}
For $N\rightarrow\infty$ the deviation is zero as expected. We take shells with width $\Delta r$ such that the difference between the free-fall times at the shell boundaries is several times the Keplerian orbit time around the newly formed compact object.
This ensures that during the collapse, a disk has time to form in case the specific angular momentum of the infalling matter is sufficient.
To do so we need to derive the change in free-fall time with radius:
\begin{equation}
\frac{dt_{ff}}{dr}=2^{-\nicefrac{5}{2}}\frac{dt_{\Kep}}{dr}=2^{-\nicefrac{5}{2}}\frac{d}{dr}\frac{2\pi r^{\nicefrac{3}{2}}}{\sqrt{GM(r)}}=\frac{\pi}{2^{\nicefrac{5}{2}}}\sqrt{\frac{r}{GM(r)}}\left(3- \frac{d \ln M}{d \ln r}\right).
\label{eqdtkep}
\end{equation}
The effect of rotation on the free-fall time is not taken into account, as we assume negligible rotation throughout the paper. The interplay between stochastic and global angular momentum will be studied in a forthcoming paper.

We now take $\Delta t_{ff}\left(r\right)$ to be several times, denoted $n_{\Kep}$,
the Keplerian time on the NS surface or black hole (BH) last stable orbit,
and derive the width of a shell that can form a temporary accretion disk
\begin{equation}
\Delta r\simeq\frac{2^{\nicefrac{5}{2}}}{\pi}\sqrt{\frac{GM(r)}{r}}\left(3- \frac{d \ln M}{d \ln r}\right)^{-1}n_{\Kep}t_{{\Kep},co}.
\label{eqdelr}
\end{equation}
In the calculations presented in this paper we take, somewhat arbitrarily, $n_{\Kep}=3$.
Using the Keplerian orbital time
\begin{equation}
t_{{\Kep},co}=\frac{2\pi {R_{co}}^\nicefrac{3}{2}}{\sqrt{GM_{co}}},
\label{eqtkepgr}
\end{equation}
where $M_{co}$ denotes the mass of the compact object and $R_{co}$ is its radius (or last stable orbit, for a BH), we derive
\begin{equation}
\Delta r\simeq 2^{\nicefrac{7}{2}}n_{\Kep}\sqrt{\frac{\nicefrac{M(r)}{r}}{\nicefrac{M_{co}}{R_{co}}}}\left(3- \frac{d \ln M}{d \ln r}\right)^{-1}R_{co}.
\label{eqtkep}
\end{equation}
Then, assuming each element is a sphere of size $a_{c}$ (which can be taken to be the mixing length, for example), the number of elements in a shell is:
\begin{equation}
N=\frac{V_\mathrm{shell}}{V_\mathrm{element}}=\frac{4\pi r^{2}\Delta r}{\frac{4\pi a_{c}^{3}}{3}}=\frac{3r^{2}\Delta r}{a_{c}^{3}}.
\label{eqn}
\end{equation}
Substituting (\ref{eqtkep}) into (\ref{eqn}), and (\ref{eqn}) into (\ref{eqsigmaj}), we get:
\begin{equation}
\sigma(j_{z})=\frac{2^{-\nicefrac{5}{4}}}{3^\nicefrac{3}{2}n_\Kep^\nicefrac{1}{2}}
\left(\frac{\nicefrac{M_{co}}{R_{co}}}{\nicefrac{M(r)}{r}}\right)^\nicefrac{1}{4}
\left(\frac{a_c}{R_{co}}\right)^\nicefrac{1}{2}
\left(3- \frac{d \ln M}{d \ln r}\right)^\nicefrac{1}{2}
{a_c}{v_c}.
\label{eqsigmajz2}
\end{equation}

The minimal specific angular momentum required in order not to fall into the compact object is
\begin{equation}
j_{co}=\nicefrac{\sqrt{12}GM_{co}}{c},
\label{jmin}
\end{equation}
which is also applicable for a NS, since
\begin{equation}
R_\mathrm{lso}=3R_s=12.5\left(\frac{M_{co}}{1.4M_\odot}\right)\km.
\label{eqrlso}
\end{equation}
Finally, the ratio between the stochastic specific angular momentum at $r$ and the minimal specific angular momentum required not to fall into the compact object is:
\begin{equation}
\frac{\sigma(j_{z})}{j_{co}}=\frac{2^{-\nicefrac{5}{4}}}{3^\nicefrac{3}{2}n_\Kep^\nicefrac{1}{2}}
\left(\frac{\nicefrac{M_{co}}{R_{co}}}{\nicefrac{M(r)}{r}}\right)^\nicefrac{1}{4}
\left(\frac{a_c}{R_{co}}\right)^\nicefrac{1}{2}
\left(3- \frac{d \ln M}{d \ln r}\right)^\nicefrac{1}{2}
\frac{{a_c}{v_c}}{\nicefrac{\sqrt{12}GM_{co}}{c}}.
\label{eqratio}
\end{equation}
Our condition for accretion disk formation will be $\sigma / j_{\rm co} >1$.
The specified $z$ direction is of course arbitrary, and over time the direction of the angular momentum from fluctuations will change. The accretion disk will be intermittently destroyed and then rebuilt with a different orientation.

\section{FORMATION OF INTERMITTENT ACCRETION DISK}
\label{sec:models}

\begin{figure}
\begin{tabular}{cc}
{\includegraphics*[scale=0.41]{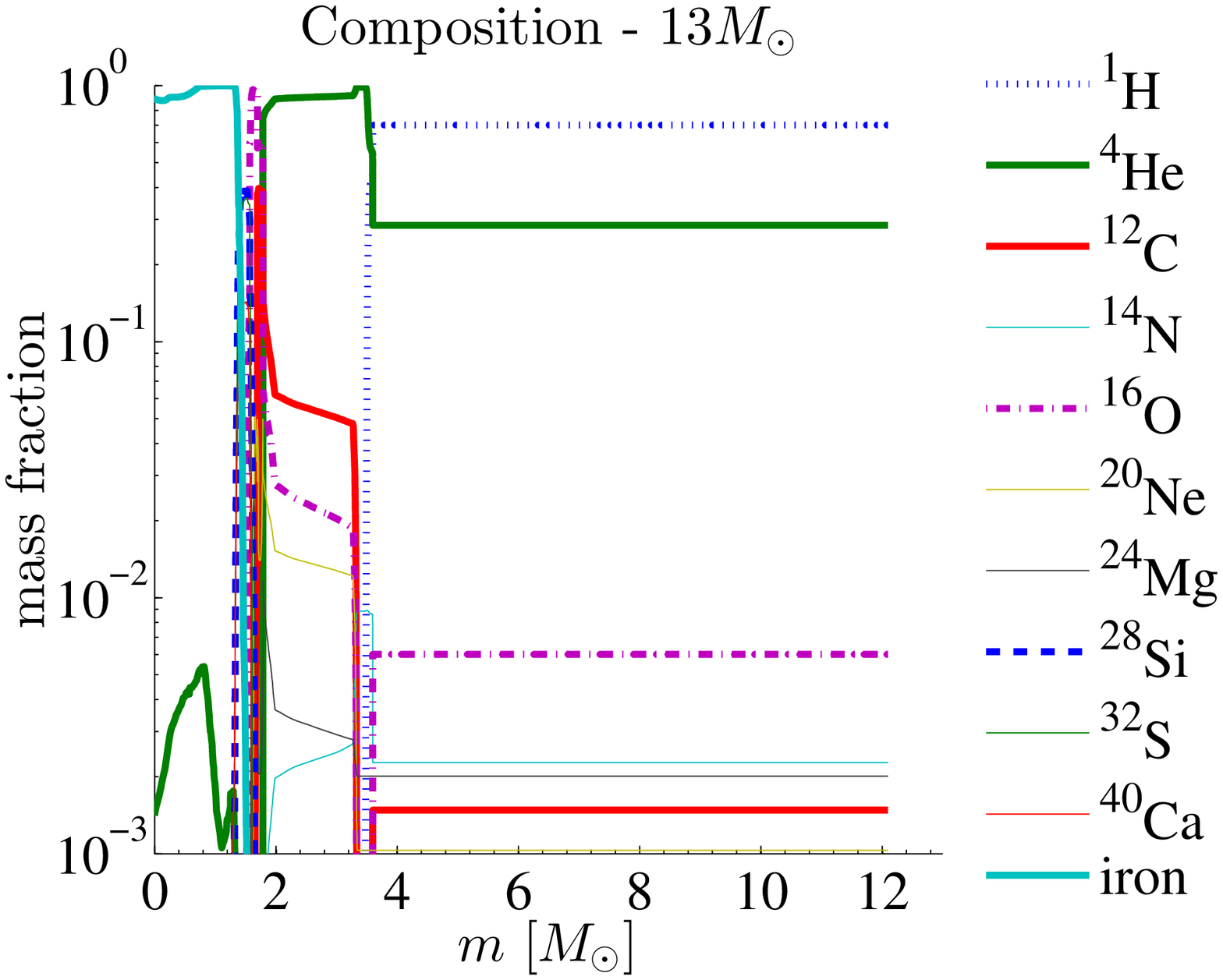}} &
{\includegraphics*[scale=0.41]{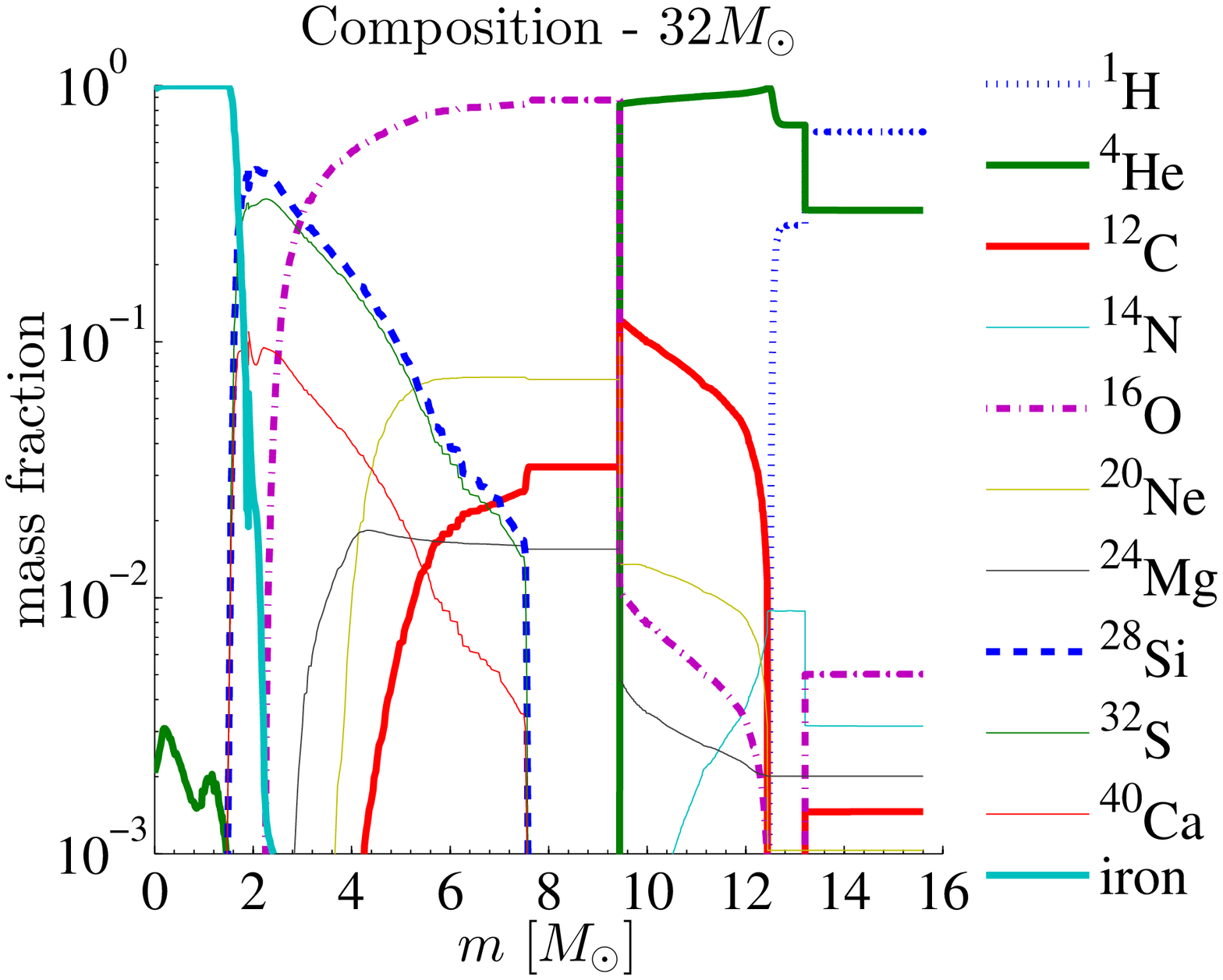}} \\
{\includegraphics*[scale=0.41]{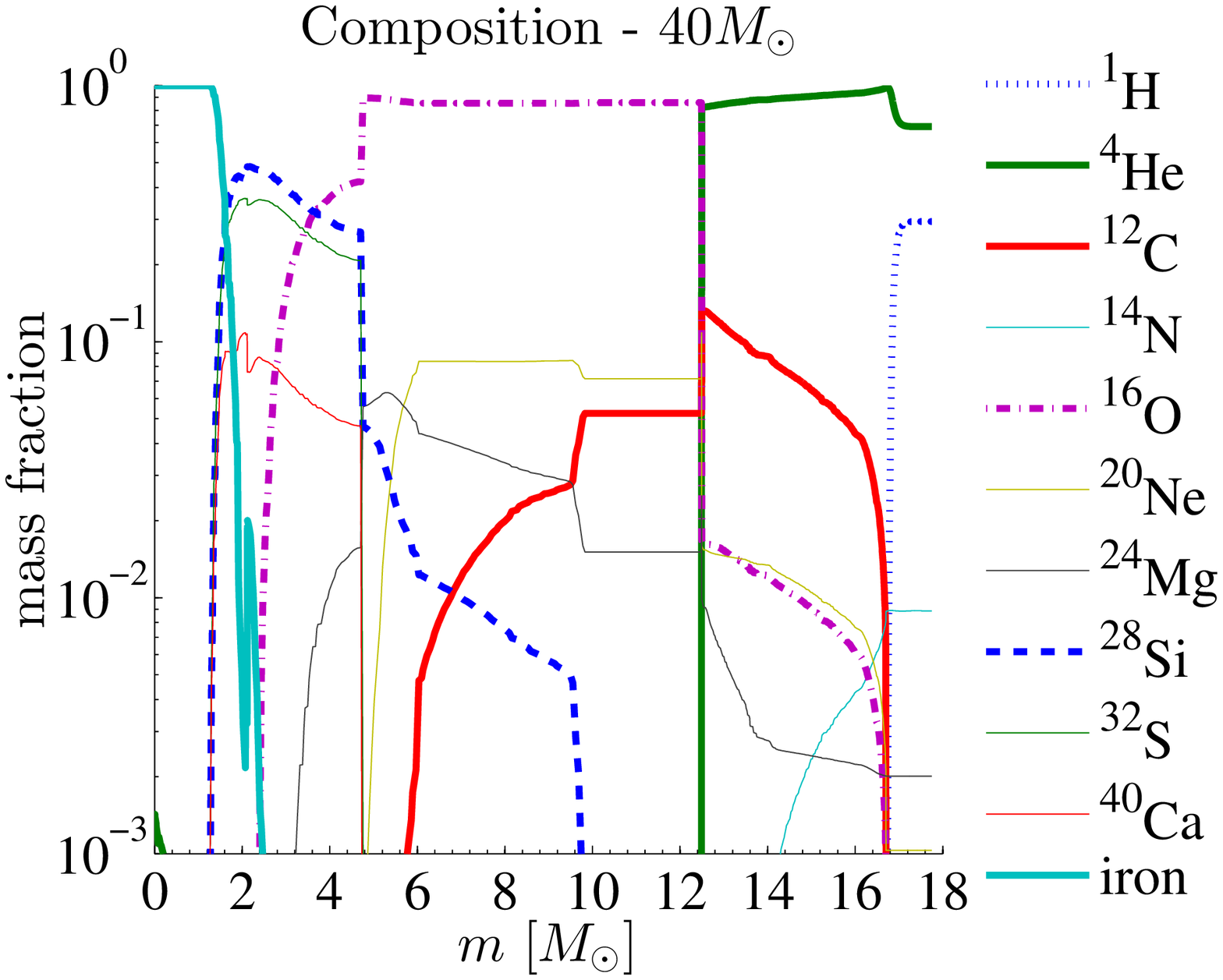}} &
{\includegraphics*[scale=0.41]{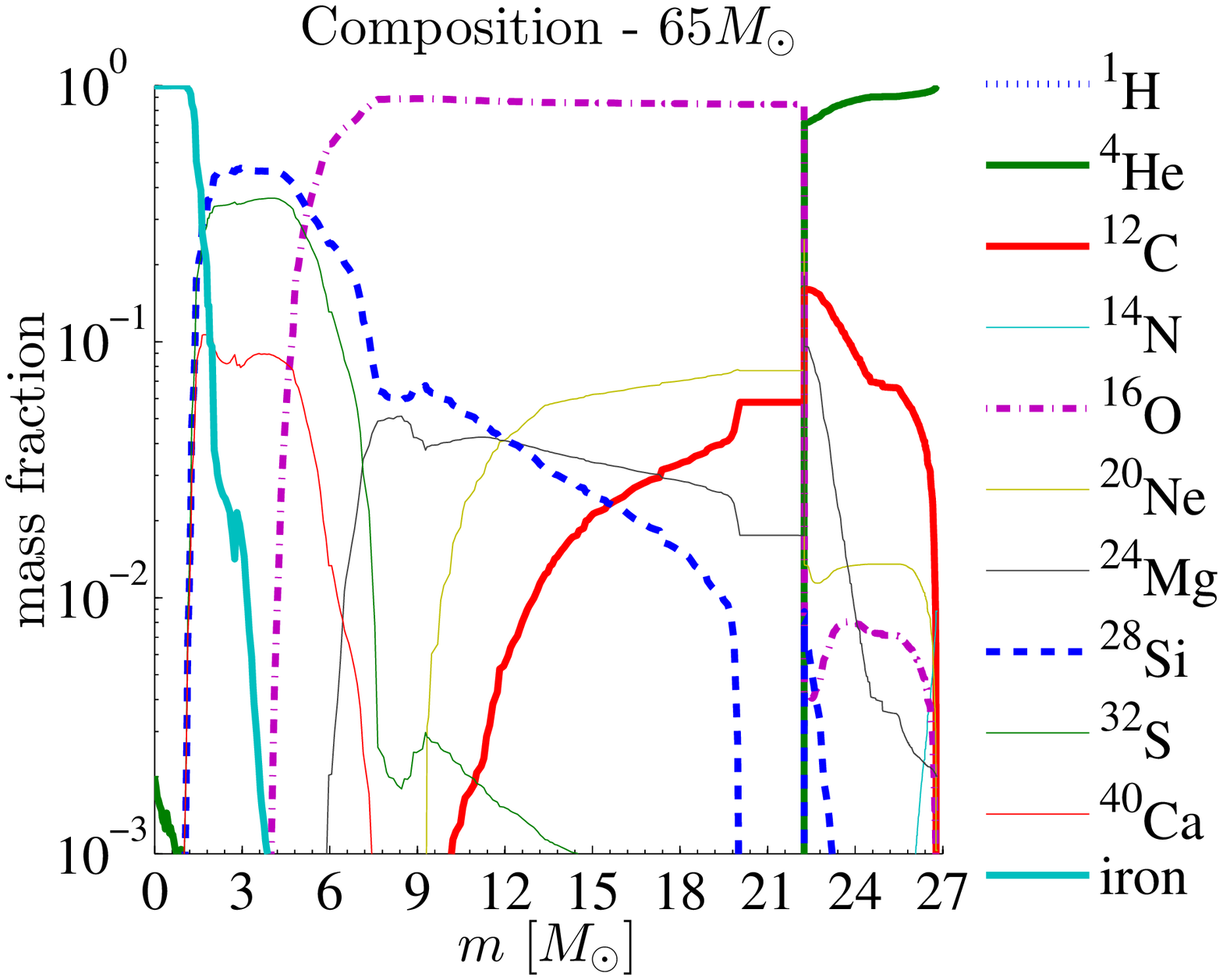}} \\
\end{tabular}
      \caption{Composition of the stellar models at the Si-shell burning stage. All models are of non-rotating stars with initial solar metallicity ($Z=0.014$). Models are marked by their initial mass.}
      \label{fig:composition}
\end{figure}

We present four stellar models constructed by Modules for Experiments in Stellar Astrophysics (MESA version 5819; \citealt{Paxton2011,Paxton2013}),
with initial masses of $M_\zams=13$, $32$, $40$ and $65 M_\odot$.
The models are of non-rotating solar metallicity stars ($Z=0.014$), and magnetic fields are neglected.
All models were evolved well into the silicon shell burning stage,
and have an iron-core mass of $\sim 1.5M_{\odot}$. Due to stellar winds
(we use the so-called 'Dutch' scheme, e.g., \citealt{Nugis2000,Vink2001})
the final masses are $12M_\odot$, $16M_\odot$, $18M_\odot$ and $27M_\odot$, respectively. The heaviest model, with $M_\zams=65M_\odot$,
loses its hydrogen envelope and becomes a Wolf-Rayet (WR) star. The three lighter models become supergiants of different temperature and
therefore color classification. The main characteristics of the models are summarized in Table \ref{table:parameters1},
and their detailed composition structures are presented in Figure \ref{fig:composition}.
\begin{deluxetable}{lcccc}
\tablecolumns{5}
\tablewidth{0pc}
\tablecaption{The model parameters at the Si-shell burning stage}
\tablehead{
    \colhead{Model} & \colhead{I} & \colhead{II} & \colhead{III} & \colhead{IV} \\
    \colhead{} & \colhead{RSG} & \colhead{YSG} & \colhead{BSG} & \colhead{WR}
}
\startdata
    $\mathrm{Initial\ mass}\;[M_\odot]$ & $13$ & $32$ & $40$ & $65$ \\
    $\mathrm{Pre-explosion\ mass}\;[M_\odot]$ & $12$ & $16$ & $18$ & $27$ \\
    $R\;[R_\odot]$ & $699$ & $892$ & $114$ & $0.66$ \\
    $L\;[L_\odot]$ & $5.19\times10^4$ & $4.63\times10^5$ & $6.12\times10^5$ & $1.23\times10^6$ \\
    $T\;[K]$ & $3,298$ & $5,102$ & $15,141$ & $237,249$ \\
	Figure & \ref{fig:model13M} & \ref{fig:model32M} & \ref{fig:model40M} & \ref{fig:model65M} \\
\enddata
\tabletypesize{\footnotesize}
\flushleft
{Mass, radius, luminosity and effective temperature of the stellar models just before explosion.
The first row mark the initial mass.
RSG, YSG and BSG stand for red super-giant, yellow super-giant, and blue super-giant, respectively.}
\\[1.5ex]
\label{table:parameters1}
\end{deluxetable}

For each model we calculate the ratio between the standard deviation of specific angular momentum and the specific angular momentum for a Keplerian orbit around the newly formed compact object, as function of the stellar radius from where the gas is accreted, according to equation (\ref{eqratio}). The convection element size is according to the mixing-length theory with $\alpha=1.5$ (default in MESA for massive stars). For the $M_\zams=13M_\odot$ model, we assume the compact object is a NS of radius $12\km$, and a baryonic mass of $2M_\odot$. This mass is chosen as the mass coordinate where strong convection begins, and according to our model jets may form and prevent further mass accretion. Due to gravitational mass loss by neutrinos, the NS final mass will be $\sim10 \%$ lower. For the heavier pre-explosion models we assume the compact object is a BH, whose mass is the total mass encapsulated within for each mass coordinate.  This too may be lower as the NS has to cool before becoming a BH, while losing mass by neutrino emission.

\begin{figure}
\begin{tabular}{cc}
{\includegraphics*[scale=0.4]{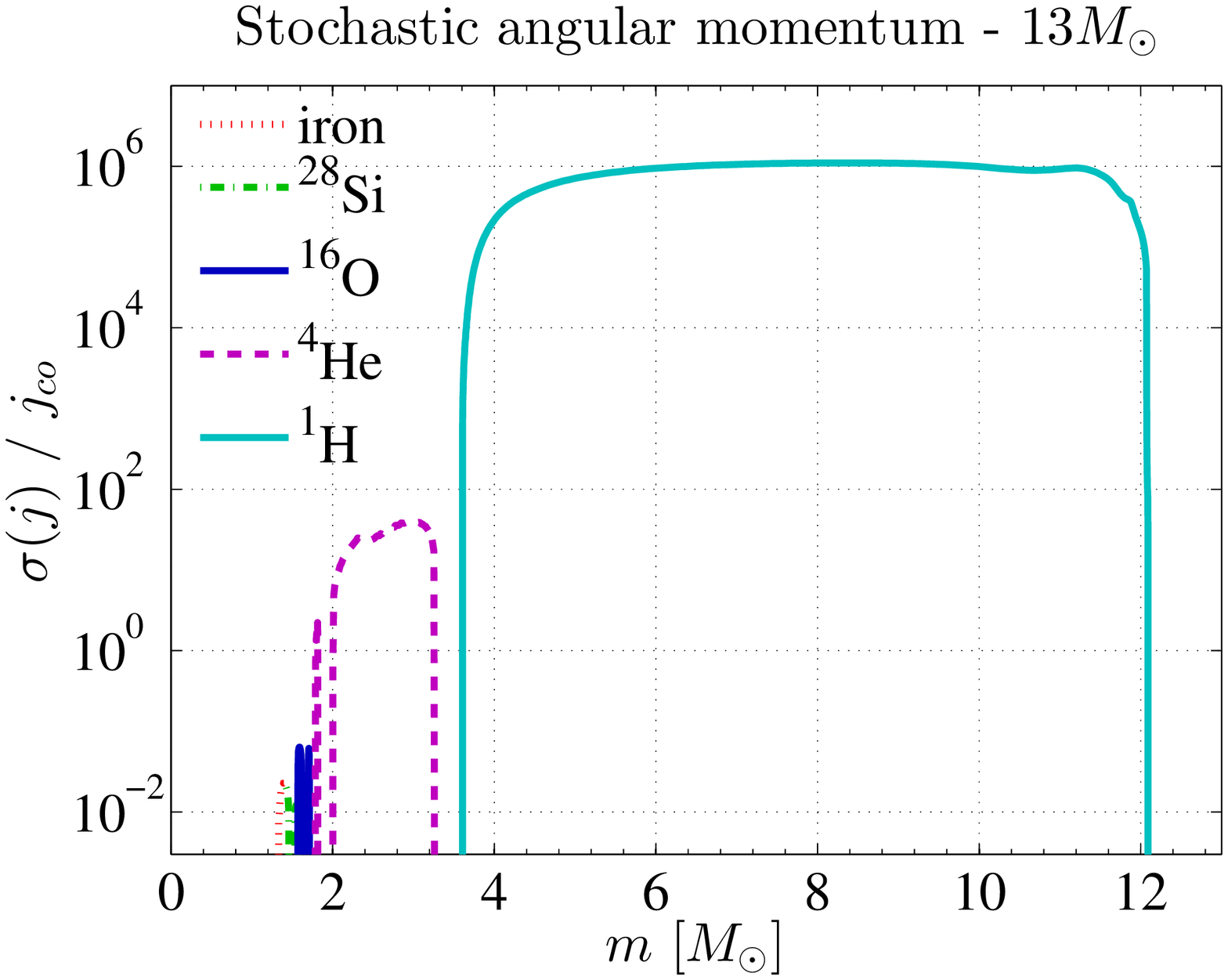}} &
{\includegraphics*[scale=0.4]{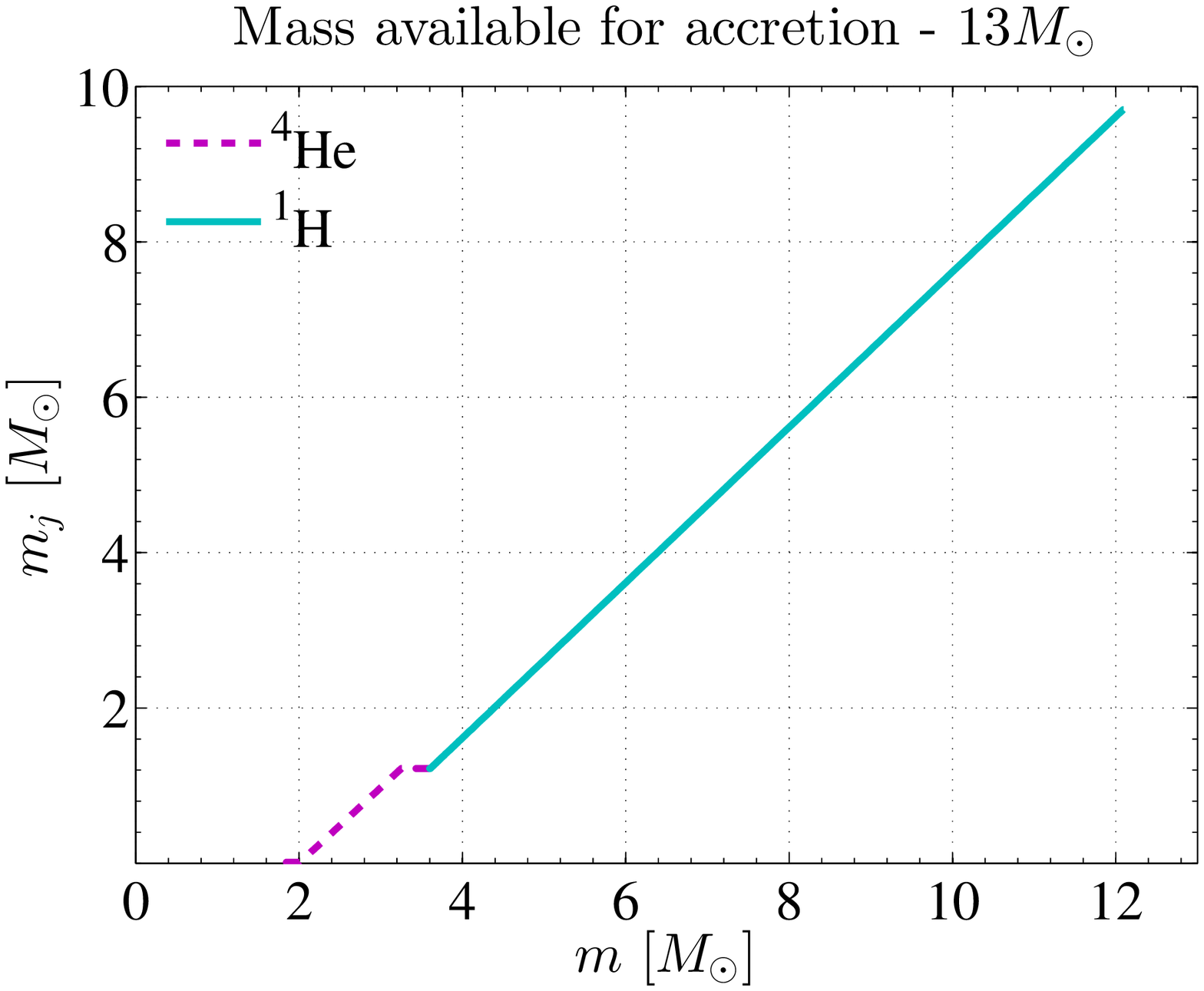}} \\
\end{tabular}
      \caption{\textit{Left:} Ratio between the standard deviation of the specific angular momentum of convective cells to the specific angular momentum of a Keplerian orbit around the newly formed NS, for the $M_\zams=13M_\odot$ model at the silicon shell burning stage. This quantity is calculated from equation (\ref{eqratio}) with $n_\Kep=3$. The different line styles represent the dominant material in each region, as indicated by the inset. \textit{Right:} Accumulated mass with high enough specific angular momentum for disk formation. The different line styles represent the dominant material in each convective region that supplies that gas, as indicated by the inset. According to \cite{Papish2014a}, accretion of $0.06-0.1M_\odot$ onto a NS is enough to explode the star; here  accretion of $0.1M_\odot$ is attained at a mass coordinate of $m=2.1M_\odot$.}
      \label{fig:model13M}
\end{figure}

The $M_\zams=13M_\odot$ case is presented in Figure \ref{fig:model13M}. In the left panel of Figure \ref{fig:model13M}
it can be seen that the hydrogen envelope has large deviations in its angular momentum.
This is chiefly due to a large pressure scale height which implies sizeable convection elements,
and the large radii in which the hydrogen is situated
(see eq. \ref{eqratio}). However, the hydrogen envelope might not be relevant for the formation of an accretion disk,
if: ($i$) the more inner parts succeed in creating jets which expel the outer parts; ($ii$) the hydrogen envelope
is ejected due to the decrease in gravitational force as the newly born NS losses $\sim {\rm few} \times 0.1 M_\odot$
to neutrinos \citep{Nadezhin1980, Lovegrove2013}. If neither scenario happens, then accretion of hydrogen will ensue,
forming jets and an energetic transient event. In this case the remnant object will be a BH, and not a NS, as the
convective hydrogen envelope only starts at a mass coordinate $m=3.6M_\odot$. To prevent BH formation the convective
regions of the core must fuel the production of intermittent accretion disks.

The helium shell has the most prominent angular momentum deviations, after the hydrogen envelope.
Quantitatively, we can assume a Gaussian distribution
of the specific angular momentum around the mean value
and estimate the mass fraction which has more angular momentum than needed for a Keplerian orbit
(eq. \ref{eqratio} with $n_{\rm Kep}=3$).
Integration yields an accumulating mass which is available for accretion, as presented in the right panel of Figure \ref{fig:model13M}.
The unity slope of the mass is due to most of the material in the hydrogen and helium regions
having enough specific angular momentum to form an accretion disk.
The total mass contribution of the helium shell is $\sim1.2M_{\sun}$.
\textit{This is enough mass by a large margin for the formation of jets with sufficient energy to attain a successful supernova explosion.}

We ran stellar models with initial masses of $15 M_\odot$ and $25 M_\odot$,
as done by \cite{Lovegrove2013}, and reached the same conclusion.
Namely, there is sufficient mass in the helium layer to form intermittent accretion disks to power a supernova.
We conclude therefore that a very weak explosion with an energy of $\la 10^{48} \erg$,
as proposed by \cite{Nadezhin1980} and \cite{Lovegrove2013}, will not take place as jets will lead to a much more powerful explosion.

Over many epochs of the stochastic accretion process the angular momentum of the accreted mass changes direction by a large value.
During these transition periods the angular momentum might sum-up to a small specific angular momentum insufficient for disk formation.
So even if each convective element has sufficient specific angular momentum
to form a disk, the total mass that will be accreted through an accretion disk is less than the mass that
we mark available for disk formation (e.g. right panel of Figure \ref{fig:model13M}).
On the other hand, directional accretion \citep{Papish2014b}, slow core rotation,
and instabilities in the infalling gas, all to be studied in forthcoming papers, are likely to
increase the amount of gas available for disk formation.

\begin{figure}
\begin{tabular}{cc}
{\includegraphics*[scale=0.4]{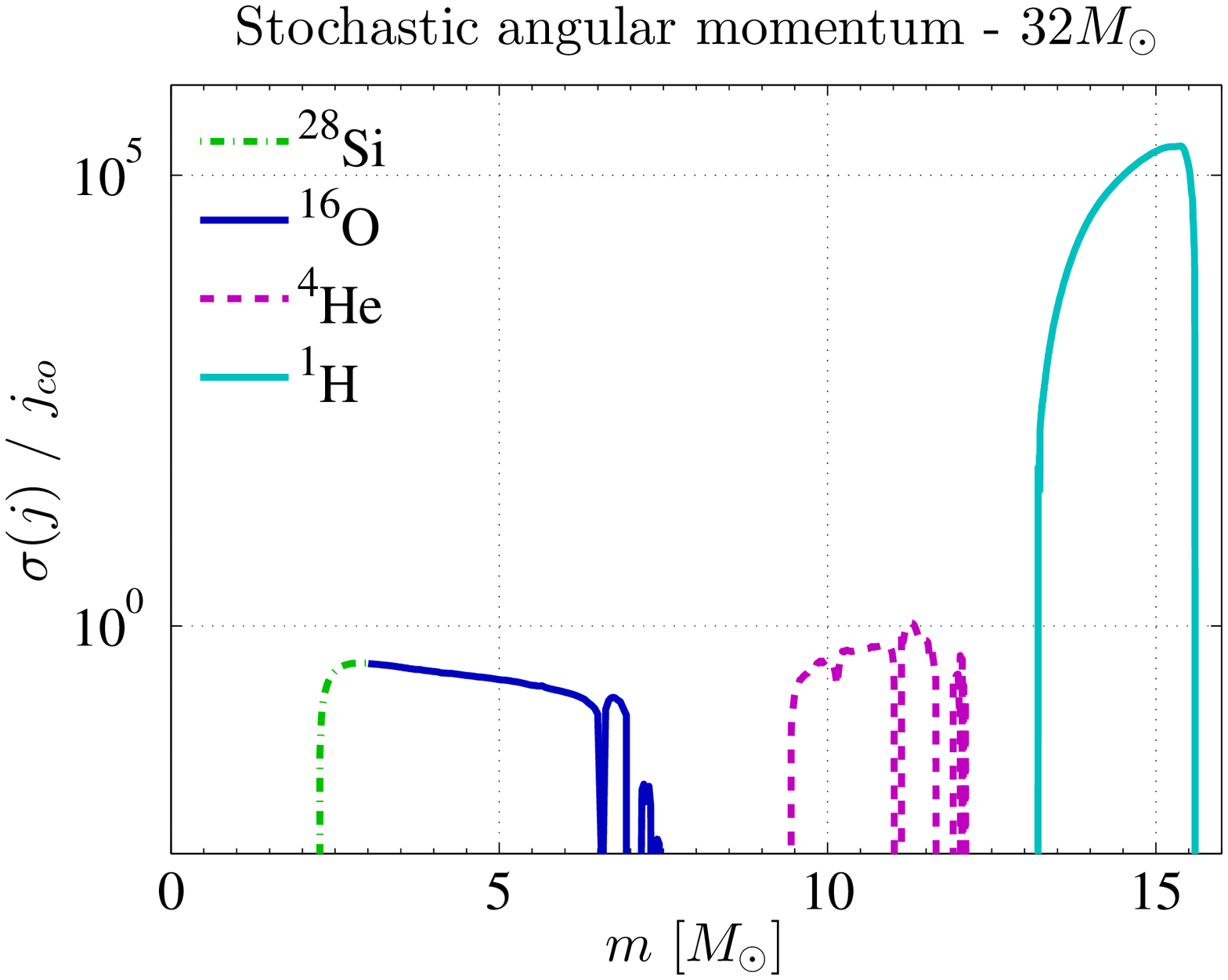}} &
{\includegraphics*[scale=0.4]{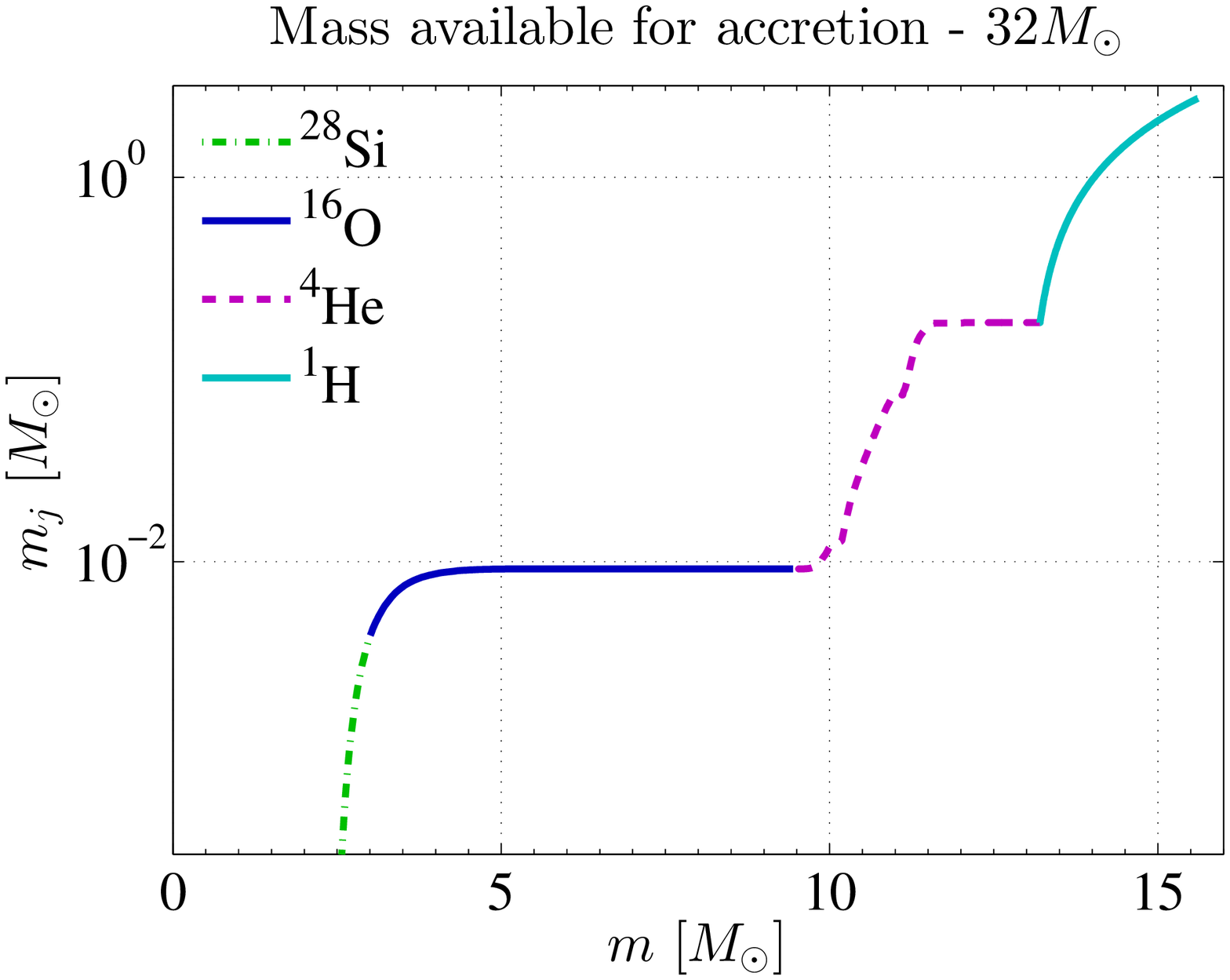}} \\
\end{tabular}
      \caption{Like Figure \ref{fig:model13M}, but for the $M_\zams=32M_\odot$ model, and accretion around a BH rather than a NS. Accretion of $0.1M_\odot$ is attained at a mass coordinate of $m=11.2M_\odot$.
      A BH can be more efficient than a NS in powering jets, and the jets might explode the star before all the available mass is accreted.}
      \label{fig:model32M}
\end{figure}

The $M_\zams=32M_\odot$ model is presented in Figure \ref{fig:model32M}. The left panel of Figure \ref{fig:model32M} shows that strong convection starts around $m=2.3M_\odot$, at the oxygen burning shell. However, the deviations in stochastic specific angular  momentum are relatively small, and it is unclear whether jets will form and halt further accretion. Uncertainties in the maximal possible NS mass together with the relatively low mass available for accretion ($\sim0.01M_\odot$ from the oxygen shell) make it difficult to predict whether a NS forms or a BH.
Let us assume indeed that a BH forms, and then $0.01 M_\odot$ from the oxygen layer is accreted and forms an intermittent accretion disk.
Taking the canonical efficiency of transforming rest mass to jet kinetic energy of $f_\mathrm{BH} \sim 0.1$, we find the energy in the jets to be 
$\sim 10^{-3} M_\sun c^2 = 2 \times 10^{51} \erg$ - this is a SN explosion.

As in the $M_\zams=13M_\odot$ case, the angular momentum deviations in the hydrogen are such that a large fraction of the envelope mass may form an 
intermittent accretion disk. Again, this mass may not be relevant for accretion, for the reasons detailed previously.
After the hydrogen envelope, the helium shell, starting at $m=9.44M_\odot$, has the most considerable deviations of angular momentum.
Taking an efficiency of $f_\mathrm{BH}=0.1$ we find that $0.1 M_\odot$ accreted through a disk can lead to jets with $\sim 10 ^{52} \erg$.
Even an efficiency of $f_{\rm BH}=0.01$ can lead to a typical SN explosion energy.
Although a mass of $0.175M_\odot$ is available for accretion, once jets start to be launched they suppress or even stop altogether
further accretion via a negative feedback mechanism.
Therefore, the value of $\sim 10^{52} \erg$ will not be reached in many cases.
Only when the negative feedback mechanism \emph{is less efficient we will get a very energetic explosion}.

\begin{figure}
\begin{tabular}{cc}
{\includegraphics*[scale=0.4]{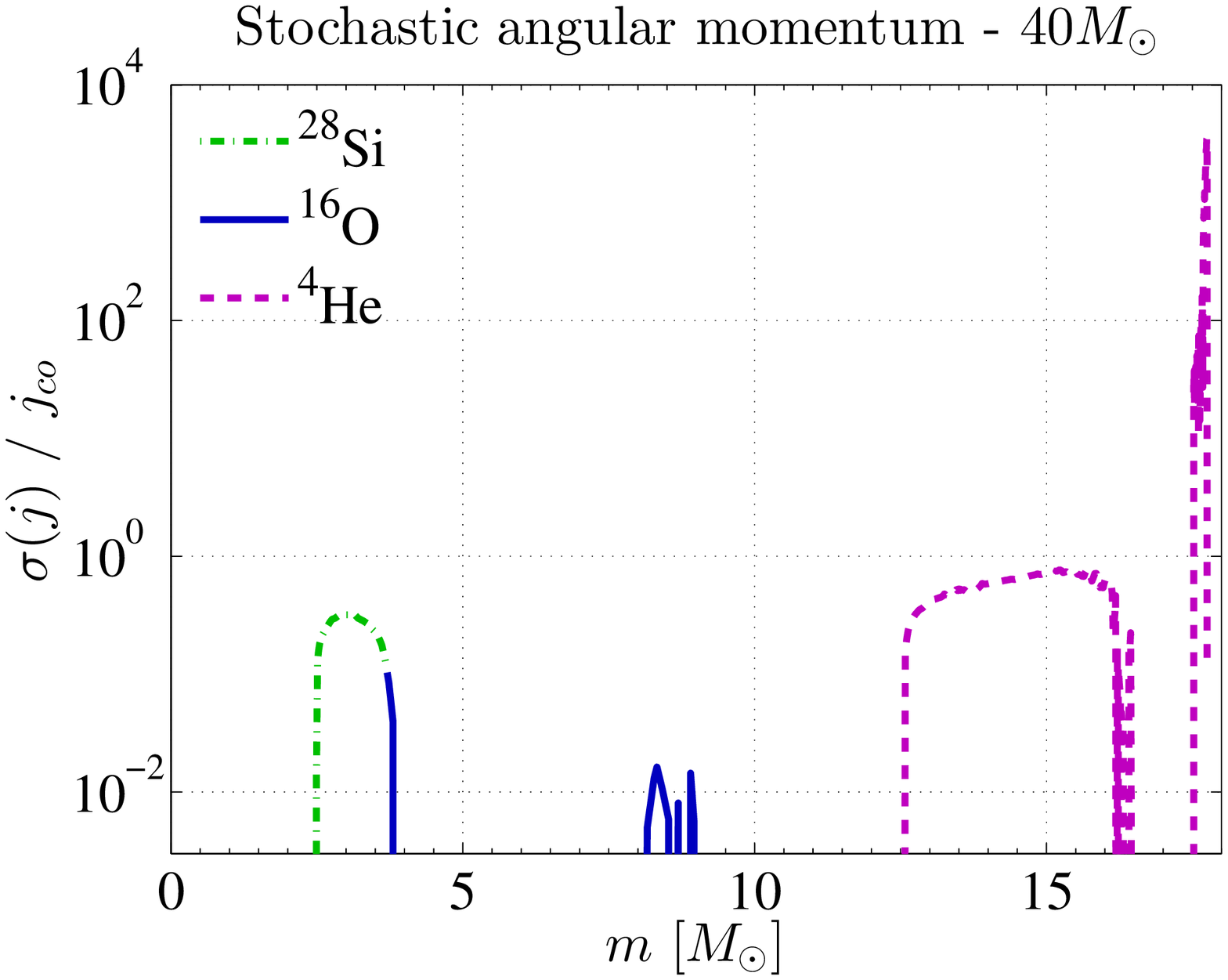}} &
{\includegraphics*[scale=0.4]{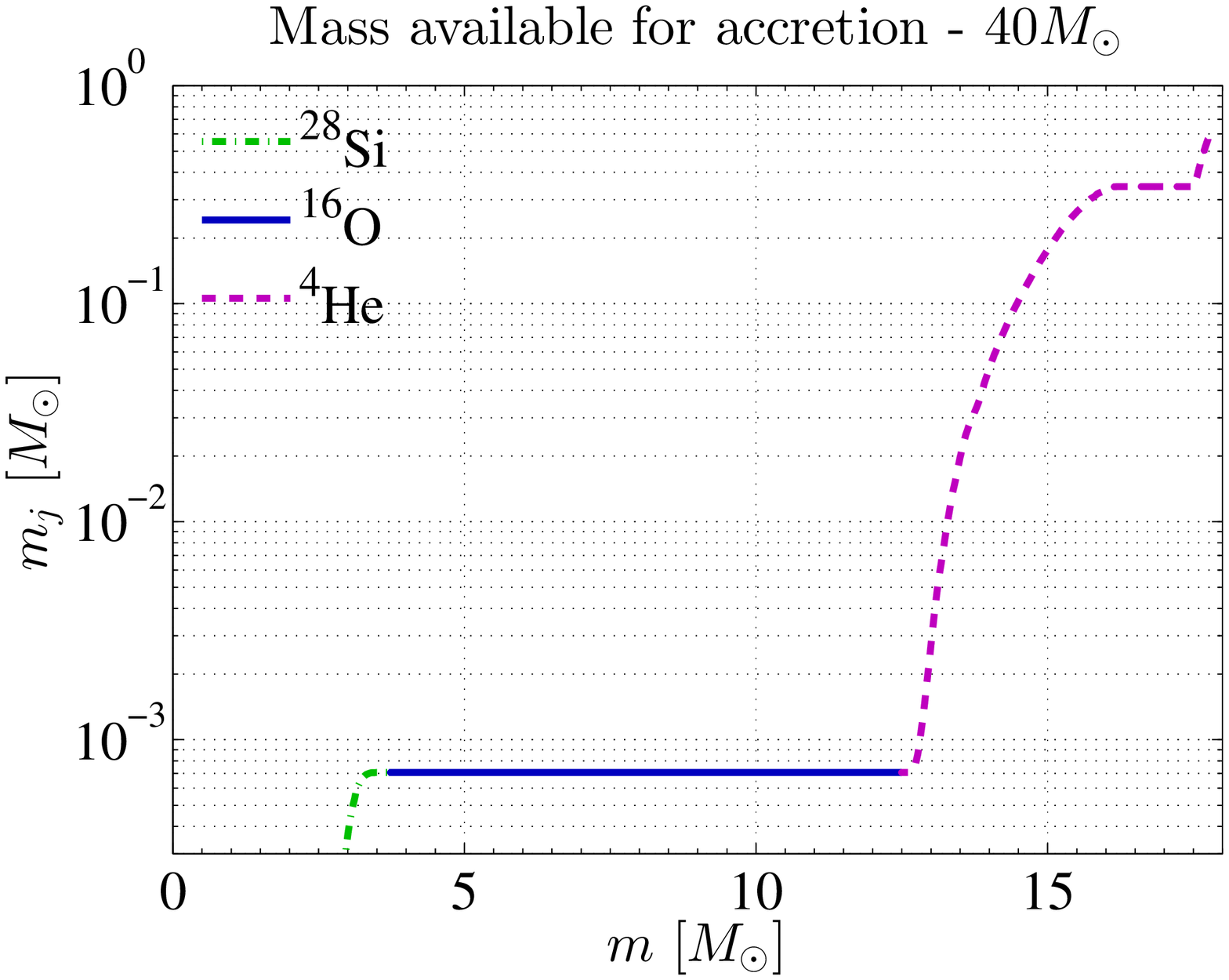}} \\
\end{tabular}
      \caption{Like Figure \ref{fig:model13M}, but for the $M_\zams=40M_\odot$ model, and accretion around a BH rather than a NS. Accretion of $0.1M_\odot$ is attained at a mass of coordinate $m=14.5M_\odot$. The outermost helium envelope, between $m=17M_{\sun}$ and $m=18M_{\sun}$, contains $\sim 30 \%$ of hydrogen.}
      \label{fig:model40M}
\end{figure}

Figure \ref{fig:model40M} shows the results for the $M_\zams=40M_\odot$ model.
Strong convection starts in the silicon region at $m=2.5M_\odot$ but the estimated mass available for accretion is small  
($\sim7\times10^{-4}M_\odot$) and we assume that a BH forms.
We emphasize once more that the assumption of BH formation holds as long as there are no other sources (stochastic or not) of angular momentum.
Other sources of angular momentum, instabilities in the shocked infalling gas and pre-explosion core rotation, can lead to jets-launching earlier in the collapse process and prevent the formation of a BH. These are the subject of a future paper. In the helium burning shell, starting at $m=12.5M_{\sun}$, the stochastic deviation in angular momentum is significant. Excluding the extended helium/hydrogen envelope, this shell contributes $0.34M_{\sun}$ of mass (Figure \ref{fig:model40M}, right panel) for possible accretion-disk formation. This is enough mass for accretion that will bring about a successful jet-driven supernova explosion, which will leave a remnant BH.

\begin{figure}
\begin{tabular}{cc}
{\includegraphics*[scale=0.4]{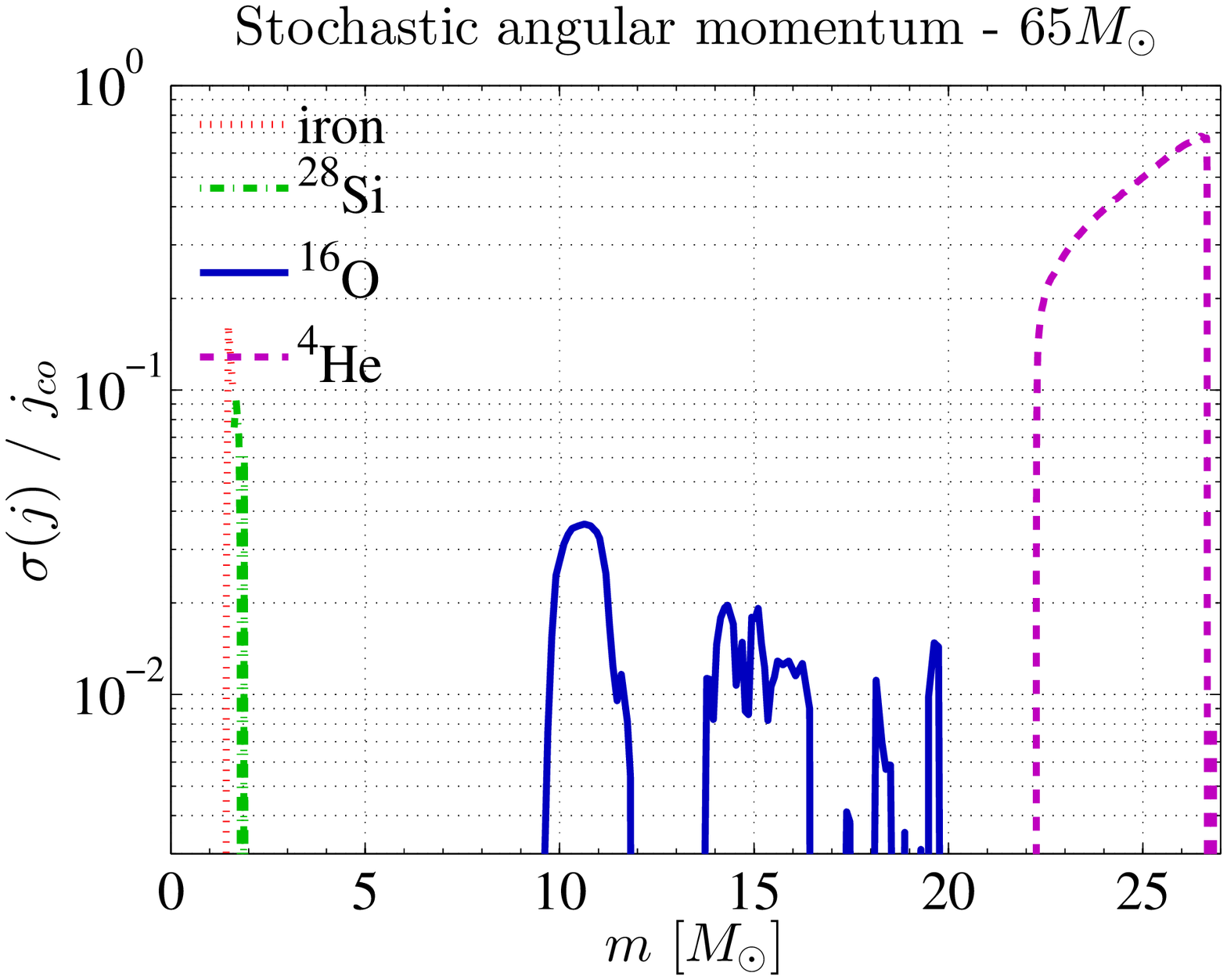}} &
{\includegraphics*[scale=0.4]{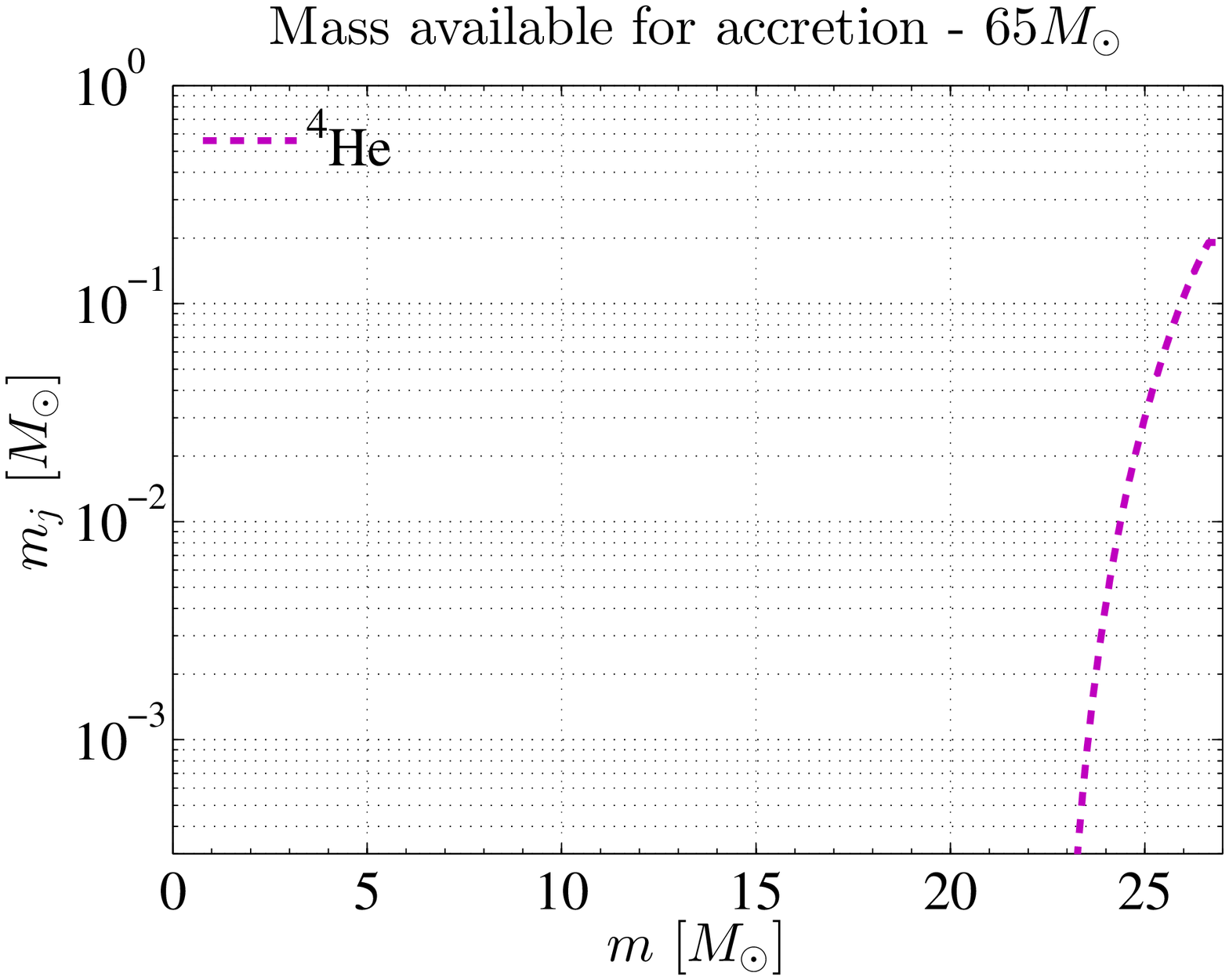}} \\
\end{tabular}
      \caption{Like Figure \ref{fig:model13M}, but for the $M_\zams=65M_\odot$ model, and accretion around a BH rather than a NS. Accretion of $0.1M_\odot$ is attained at a mass coordinate of $m=25.9M_\odot$.}
      \label{fig:model65M}
\end{figure}

Figure \ref{fig:model65M} shows the results for the $M_\zams=65M_\odot$ model. In this stellar model the hydrogen envelope has been ejected and a compact WR star remains. This star has a convective helium outer layer with non-negligible deviations in angular momentum, as shown in the left panel of Figure \ref{fig:model65M}. A mass of $\sim 0.1 M_\odot$ (right panel of Figure \ref{fig:model65M}) can fuel intermittent accretion disks and jets, possibly leading to a Type 1c supernova. It has been suggested that compact stars, such as this stellar model, are the origin of gamma-ray bursts if they have sufficient angular momentum (e.g. \citealt{MacFadyen1999}). The interplay of global angular momentum and the stochastic angular momentum presented here will be investigated in a forthcoming paper.

\section{IMPLICATIONS AND SUMMARY}
\label{sec:implicaitons}

Using the mixing-length theory we developed an approximate expression for stochastic deviation from zero angular momentum of convective mass elements in convective regions of stars. In equation (\ref{eqratio}) we derive the ratio of this specific angular momentum to the one required to form an accretion disk around the newly born neutron star (NS) or black hole (BH). The ratio given in equation (\ref{eqratio}) contains dependencies on the newly-formed compact object properties (mass and radius), as well as on the properties of convection - a typical size and velocity of convective mass elements and the location of the convection region. Although not accurate (e.g. \citealt{Arnett2011}), the mixing-length theory is adequate for the goals of the present preliminary study.

Using MESA \citep{Paxton2011,Paxton2013}, we evolved four massive stars from main-sequence masses of  $13M_\odot$, $32M_\odot$, $40M_\odot$ and $65M_\odot$, almost until core-collapse. Due to stellar winds the pre-explosion masses were $12M_\odot$, $16M_\odot$, $18M_\odot$ and $27M_\odot$, respectively. The three lighter models were super-giants prior to explosion, while the heaviest model ($M_\zams=65M_\odot$) became a WR star. We assumed for the $M_\zams=13M_\odot$ case that a NS forms, and for the heavier models a BH forms instead.
The remnant object for the intermediate $M_\zams=32M_\odot$ and $M_\zams=40M_\odot$ cases is somewhat ambiguous,
as strong convection begins at a mass coordinate similar to the upper limit of possible NS mass,
though there might not be enough mass for disk formation and jets launching.

Applying our derivation for the ratio of stochastic specific angular momentum of accreted gas from convective regions to that required to form an accretion disk around the newly born NS or BH, equation (\ref{eqratio}) with $n_\Kep=3$, we reach the following conclusions. ($i$) The hydrogen envelope has enough angular momentum to form a relatively long-lasting accretion, although this is perhaps irrelevant as the hydrogen detaches from the star. ($ii$) The inner regions, notably oxygen and more so helium, have sufficient angular momentum to form a jet-driven supernova explosion.

We studied several cases of massive stellar models where a BH most likely forms upon core-collapse. We show that the regions surrounding the newly-formed BH have large angular momentum deviations. A large enough fraction of the mass can be accreted onto the BH and form intermittent accretion disks, which in turn may generate powerful jets. Such jets can facilitate a supernova explosion, leaving behind a black hole. The explosion time-scale is short relative to the dynamic response of the envelope to the loss of gravitational mass. In this case there will be no very low energy supernova such as suggested by \cite{Nadezhin1980} and \cite{Lovegrove2013}. We instead argue that if the inner Si-rich region of the core does not manage to form intermittent accretion disks, then the helium (and in some cases the oxygen) convective region will. As the helium convective region is more massive than the Si-rich region the jets launched by these disks will carry more energy, and in the regime of the jittering-jets explosion mechanism the explosion will be stronger. {\it Namely, we argue that the failure of the inner $\sim 2-2.5 M_\odot$ of the core to explode the star will lead to a more violent supernova explosion, rather than a very low energy supernova.}

The stochastic deviation in angular momentum was derived assuming a spherically symmetric accretion. This symmetry might be broken by the standing accretion shock instability (e.g., \citealt{Hanke2013}), or by the first intermittent jet episode \citep{Papish2014b}. The study of asymmetric stochastic accretion is the next logical step for our proposed explosion mechanism. Additional avenues of investigation are the possibility of stellar rotation, and more realistic modelling of convection, perhaps using hydrodynamical simulations.

We thank an anonymous referee for very helpful comments.
This research was supported by the Asher Fund for Space Research at the Technion and the US-Israel Binational Science Foundation.


\label{lastpage}

\end{document}